\documentclass[12pt,a4paper]{article}
\usepackage{graphicx,amsmath,amsfonts}
\def \ba{\begin{eqnarray}}\def\ea{\end{eqnarray}}
\def\bc{\begin{center}}\def\ec{\end{center}}
\def\nn{\nonumber\\}
\begin{document}
\title{Formation of $\pi\mu$ atoms in $K_{\mu 4}$ decay}
\author{ S.~R.~Gevorkyan\footnote{On leave of absence from
Yerevan Physics Institute},  A.~V.~Tarasov, O.~Voskresenskaya\footnote{On leave of absence
from Siberian Physical Technical Institute}}
\maketitle
\bc Joint Institute for Nuclear Research, 141980, Dubna, Russia \ec

\begin{abstract}
We derive   the decay rate of  $\pi\mu$  atom formation in   $K_{\mu 4}$ decay.
Using the obtained expressions we calculate  the   decay rate of atom formation
and point out  that considered  decay  can give a  noticeable contribution  as a background to  the
 fundamental decay  $K^+\to \pi^+\nu\bar{\nu}$.
\end{abstract}
\section{Introduction}
The elementary atoms formation in particles collisions and decays can give an unique information on
strong interaction dynamics. The determination  of the pionium atom lifetime~\cite{dirac} allows one to
get information on  $\pi\pi$  scattering lengths, whose knowledge are crucial for verification of Chiral
Perturbation Theory predictions~\cite{colangelo01}. The accuracy of scattering lengths determination from nonleptonic decays~\cite{batley09} $K^\pm\to \pi^\pm\pi^0\pi^0$    also depends on effects caused by  the possibility of $\pi\pi$ bound state formation~\cite{gevorkyan07,gevorkyan08}. The creation of positronium atoms in $\pi^0$   Dalitz decay~\cite{afanasyev90} or its photoproduction on extended target~\cite{nemenov81,gevorkyan02}can  give  information on dependence of   interaction on the spin state of the system and on mechanism of bound state formation.\\
Since the basic works  of  Nemenov~\cite{nemenov73} which stimulated the search of elementary atoms,
the  $\pi\mu$ atom has been discovered~\cite{coombes76,aronson86} in the decays of  neutral kaons  $K_L\to \pi^+\mu^-\nu$ .\\
 In the  present work we  point out the importance  of investigation of  the  $\pi\mu$ atom formation in the decay
 \ba
 K^+\to \pi^++\pi^-+\mu^++\nu
 \ea
  ($K_{\mu 4}$ decay). The reason for this looks as follows.
 In the last years the great efforts have been done~\cite{ckm} for experimental study of  the rare decay  $K^+\to\pi^+\nu\bar{\nu}$  with the major goal of determination of  the value of   $V_{st}$ , which  is unequally predicted by the theory~\cite{buchalla99, isidori03, buras08}. At present  the experiment NA42 ~\cite{na62}  at CERN SPS  is accepted , which plans  to collect $\approx 80$ events of this rare decay\footnote{At the moment the six events are reported by CKM collaboration~\cite{ckm}.}.
Later on we calculate the probability  of   $\pi\mu$  atom formation in the $K_{\mu4}$ decay and  show  that branching rate of atom formation  is not much smaller  than the branching ratio of  fundamental process $K^+\to\pi^+\nu\bar{\nu}$. As a result  the process of  $\pi\mu$ atom formation can give a certain  contribution as a background to the  basic decay  $K^+\to\pi^+\nu\bar{\nu}$  in the relevant  kinematical regions of experiment NA62.
\section{The decay rate of  the  $\pi\mu$ atom formation}
To obtain the decay rate of the $\pi\mu$  atom   formation in $K_{\mu 4}$ decay
\ba
 K^+\to \pi^++A_{\pi\mu}+\nu
\ea
we begin from   the well known ~\cite{cabibbo65, pais68} matrix element of  the decay  (1)   written in the form of   the product of the lepton   and  hadron   currents
\ba
M=\frac{G_F}{\sqrt{2}}V_{us}^* j_\lambda J^\lambda=
\frac{G_F}{\sqrt{2}}V_{us}^*\bar{u}(k_1)\gamma_\lambda(1-\gamma_5)v(k_2)(V^\lambda-A^\lambda)
\ea
where the axial $A^\lambda$ and vector $V^\lambda$  hadronic currents:
\ba
A^\lambda&=&-\frac{i}{m_K}\left((p_1+p_2)^\lambda F+(p_1-p_2)^\lambda G+(k_1+k_2)^\lambda R\right); \nn
V^\lambda&=&-\frac{H}{m_K^3}\epsilon^{\lambda\nu\rho\sigma}(k_1+k_2)_\nu (p_1+p_2)_\rho (p_1-p_2)_\sigma
\ea
 Here and later on  $k, p_1, p_2, k_1, k_2$  are the invariant  momenta  of kaon,  pions,  muon and neutrino;
  $m_K, m_\pi, m_\mu $ are  the relevant   masses.\\
Confining   as usually  by  s and p waves  and assuming the same p-wave phase  $\delta_p$  for different  form factors one has
\ba
F=F_se^{i\delta_s}+F_pe^{i\delta_p};
\qquad
G=G_pe^{i\delta_p};
\qquad
H=H_pe^{i\delta_p}.
\qquad
R=R_pe^{i\delta_p}
\ea
The main goal   of experimental investigation~\cite{rosselet77, pislak03, batley08} is  measurements of  the
quantities $F_s, F_p, G_p, H_p,  R_p,  \delta=\delta_s-\delta_p$  as a function of three
 invariant combinations of  pions and leptons momenta $s_\pi=(p_1+p_2)^2,  s_l=(k_1+k_2)^2$
 and  $\Delta=-k(p_1+p_2)$ ~\cite{pais68}.\\
From the other hand to make up  the  $\pi\mu$ atom in the decay  (1) the negative  pion and muon should have  a similar velocities. For such  kinematic only two variables are at work, which we  choose  as   $s_\pi,  s_l$.\\
Since the binding energy of the ground state of  $\pi\mu$ atom  is small~\cite{staffin77} $\varepsilon= 1.6KeV$  the atom is a nonrelativistic system. According  to the general rules of  quantum mechanics,  the amplitude of  the  decay  (2)   can be written as the product of  the  matrix element  of the decay  (1)  taken at  equal velocities of muon and negative pion  and the square of the Coulomb wave function at the origin
\ba
M(K^+\to \pi^+A_{\pi\mu}\nu)= \frac{\Psi (r=0)}{\sqrt{2 \mu}}
M( K^+\to \pi^+\pi^-\mu^+\nu)_{v_\pi=v_\mu}
 \ea
The square of the Coulomb wave function evaluated at the origin  summed over  principal quantum  number ~\cite{aronson86}
\ba
\mid\Psi(r=0)\mid^2=\sum_{n=1}\mid\Psi_n(r=0)\mid^2=\frac{1.2}{\pi}(\alpha \mu)^3
\ea
 with $\alpha=\frac{1}{137}$  the fine structure constant and $\mu=\frac{m_{\pi}m_{\mu}}{m_{\pi}+m_{\mu}}$    reduced mass.\\
Using the well known rules, we obtain  for the  decay  rate  of  (2)
 \ba
 \Gamma=\frac{1}{(4\pi)^3m_\pi m_K}\mid\Psi(r=0)\mid^2
 \int\mid M( K^+\to \pi^+\pi^-\mu^+\nu)_{v_\pi=v_\mu}\mid^2dE_\nu dE_\pi
 \ea
 Integrations  in this expression  are going over neutrino $E_\nu$  and positive pion  $E_\pi$ energies.\\
 To calculate the square of matrix element in (8) we take advantage of the fact  that the bilinear form of lepton current
  $t_{\alpha\beta}=j_{\alpha}j_{\beta}^+$ can be written in the well  known  form (see e.g. ~\cite{okun65})
 \ba
 t^{\alpha\beta}=8\left(k_1^\alpha k_2^\beta + k_2^\alpha k_1^\beta-(k_1 k_2)g^{\alpha\beta}
 + i\epsilon^{\alpha\beta\rho\sigma} k_1^\alpha k_2^\beta\right)
 \ea
This expression has to be contracted  with the relevant form of hadronic current  $T_{\alpha\beta}$ .
As an example let us consider the convolution of  lepton tensor (9)  with the square of the first term of axial
 hadronic current in (4)
 \ba
\sum t^{\alpha\beta} T_{\alpha\beta}=\frac{8}{m_K^2}\left(2(p_1k_1+p_2k_1)(p_1k_2+p_2k_2)-
(p_1+p_2)^2(k_1k_2) \right)\mid F\mid^2
\ea
Accounting  that  muon and negative pion which  compose  atom should have the equal
velocities let us  express  their momenta  through the atom momentum   $p_a$  and  mass $ m_a$:   $p_2=\frac{m_\pi}{m_a}p_a; ~~k_1=\frac{m_\mu}{m_a}p_a $  and   introduce   the  following  Lorentz invariant combinations
\ba
q_1&=&2p_1k_2=m_K^2+m_a^2-m_\pi^2-2m_KE_a\nn
q_2&=&2p_1p_a=m_K^2-m_a^2-m_\pi^2-2m_KE_\nu \nn
q_3&=&2p_ak_2=m_K^2-m_a^2+m_\pi^2-2m_KE_\pi
\ea
As  the atom energy  in the kaon rest frame is  $E_a=m_K-E_\pi-E_\nu$  the decay  (2)   describes  by two independent variables, which in our case are the positive pion energy  $E_\pi$ and neutrino energy $E_\nu$.\\
The expression (10) can be rewritten through the above  invariants
\ba
\sum t^{\alpha\beta} T_{\alpha\beta}=\frac{4m_{\mu}}{m_am_K^2}q_1\left(q_2+2m_\pi m_a\right)\mid F\mid^2
\ea
Calculating in  the same way   all terms in the contraction of  square of  axial and vector form factors  with lepton part  we obtain for the atom formation decay rate
  \ba
 \Gamma(K^+\to \pi^+A_{\pi\mu}\nu)&=&\frac{G_F^2V_{us}^2}{m_\pi (4\pi m_K)^3}  \frac{1.2\alpha^3 \mu^3}{\pi}\int \Phi(E_\pi,E_\nu)dE_\pi dE_\nu\nn
\Phi(E_\pi,E_\nu)&=&q_1(q_2 + 2m_\pi m_a) \mid F \mid^2+q_1(q_2 - 2m_\pi.m_a)\mid G \mid^2 +
m_\nu^2q_3\mid R\mid^2\nn &+&2 (q_1q_2 - 2m_\pi^2 q_3)Re( FG^*) +
2m_\mu (m_aq_1 + m_\pi q_3)Re( FR^*) \nn
&+& 2m_\mu(m_aq_1 - m_\pi q_3) Re( RG^*) +\frac{m_\pi^2}{m_a^2}\left(4E_\nu E_\pi q_1-q_1^2-4m_\pi^2E_\nu^2\right)\nn
&\times&\left(\frac{q_3}{m_\pi^2}\mid H\mid^2-2\frac{m_a}{m_\pi}Re(GH^*+FH^*)\right)
 \ea
The integration in this expression  is going   in the limits
\ba
\frac{m_K^2+m_{\pi}^2-m_a^2-2m_KE_\pi}{2(m_K-E_\pi+\sqrt{E_\pi^2-m_\pi^2})}\leq &E_{\nu}&\leq
\frac{m_K^2+m_{\pi}^2-m_a^2-2m_KE_\pi}{2(m_K-E_\pi-\sqrt{E_\pi^2-m_{\pi}^2})}\nn
 m_{\pi}\leq &E_\pi &\leq \frac{m_K^2+m_\pi^2-m_a^2}{2m_K}
\ea
The expression (13) is the main result of the present work. It  allows one to calculate  not only the decay rate of
atom formation in $K_{\mu4}$ decay,  but  also the differential decay rate  $\frac{d\Gamma}{dE_\pi}$,
whose  knowledge  is important   for estimation of  background in the basic decay  $K^+\to \pi^+\nu\bar{\nu}$.
\section{Numerical analysis}
To  calculate   the  atom formation decay rate  using expression    (13)  one has  to  know the  hadronic form factors.
Accounting that hadronic form factors in $K_{\mu4}$   and  $K_{e4}$ decays are the same we take
for three form factors F, G  and H the standard  parametrization ~\cite{rosselet77, pislak03, batley08} with
parameters\footnote{The precision  of the experimental data~\cite{batley08} are better than in ~\cite{pislak03}, but unfortunately only relative parameters determining form factors are cited.} from~\cite{pislak03}.\\
The axial hadronic form factor R can not be  extracted    from the experimental data on  $K_{e4}$ decay
\footnote{ The term with R in $K_{e4}$  decay rate is proportional to the  square of  electron mass and  can be
neglected.}. For this quantity we use  the theoretical prediction ~\cite{knecht93} $ R=\frac{2}{3}F$.
Substituting  these parametrization   in (13)  and using the  value of  $K_{\mu4}$  decay rate  from ~\cite{pdg} we obtain for the  atom formation probability  in  the $K_{\mu4}$ decay
 $\Gamma(K^+\to \pi^++A_{\pi\mu}+\nu)/\Gamma(K^+\to \pi^++\pi^-+\mu^++\nu)\approx 3.7\times 10^{-6}$.
This probability would be compared with  the probability of  $\pi\mu$ atom creation in   $K_{\mu3}$  decay~\cite{aronson86} $\sim 4\times 10^{-7}$ and   $\pi\pi$  atom formation in the nonleptonic decay~\cite{silagadze94}   $K^+\to \pi^+\pi^+\pi^-$    $\sim 8\times 10^{-6}$.\\
As was mentioned above the atom formation in $K_{\mu4}$  decay  can serves as background to  the    rare decay $K^+\to \pi^+\nu\bar{\nu}$  in the relevant kinematical region.  The  Standard Model  predicts  for the  branching decay  rate (see e. g. ~\cite{brod08})  $Br( K^+\to \pi^+\nu\bar{\nu})\approx (0.85\pm 0.07)\times 10^{-10}$  whereas  the branching ratio for $\pi\mu$ atom formation considered in the present work  turn out to be  $Br( K^+\to \pi^+A_{\pi\mu}\nu)\approx 0.5\times 10^{-10}$.
Thus the branching ratio of  the decay (2) considered in present work  is comparable with branching ratio of basic decay $K^+\to \pi^+\nu\bar{\nu}$ and so has to be considered as a possible  background to this  decay.\footnote{The corresponding consideration will be done elsewhere.}\\
The authors  are grateful to V. Kekelidze, D. Madigozhin and Yu. Potrebenikov   for permanent support and  useful
discussions.

\end{document}